\documentclass[fleqn,usenatbib]{mnras}
\usepackage[T1]{fontenc}
\usepackage{graphicx}	% Including figure files
\usepackage{amsmath}	% Advanced maths 
\title[RLEES II. uGMRT observations of the hot-Saturn WASP 69b]{uGMRT observations of the hot-Saturn WASP 69b: Radio-Loud Exoplanet-Exomoon Survey II (RLEES II) }

\author[Narang et al.]{
Mayank Narang$^{1,2}$\thanks{E-mail: mnarang@asiaa.sinica.edu.tw},
Apurva V. Oza$^{3}$, 
Kaustubh Hakim$^{4,5}$,
P. Manoj$^{2}$, Himanshu Tyagi$^{2}$, Bihan  Banerjee$^{2}$, \newauthor Arun Surya $^{2}$, Prasanta K. Nayak $^{2}$, Ravinder K. Banyal $^{6}$, Daniel P. Thorngren$^{7}$
\\
$^{1}$ Academia Sinica Institute of Astronomy \& Astrophysics, 11F of Astro-Math Bldg., No.1, Sec. 4, Roosevelt Rd., Taipei 10617, Taiwan, R.O.C.\\
$^{2}$Tata Institute of Fundamental Research, Mumbai, India; $^{3}$Jet Propulsion Laboratory, California Institute of Technology, Pasadena, USA\\
$^{4}$
KU Leuven, Institute of Astronomy, Celestijnenlaan 200D, 3001 Leuven,
Belgium; $^{5}$Royal Observatory of Belgium, \\ Ringlaan 3, 1180 Brussels, Belgium ; $^{6}$ Indian Institute of Astrophysics, Bangalore, India; $^{7}$ Universit\'{e} de Montr\'{e}al, Quebec
}

\begin{document}

\label{firstpage}
\pagerange{\pageref{firstpage}--\pageref{lastpage}}
\maketitle

\begin{abstract}
Exomoons have so far eluded ongoing searches. Several studies have exploited transit and transit timing variations and high-resolution spectroscopy to identify potential exomoon candidates. One method of detecting and confirming these exomoons is to search for signals of planet-moon interactions. In this work, we present the first radio observations of the exomoon candidate system WASP 69b. Based on the detection of alkali metals in the transmission spectra of WASP-69b, it was deduced that the system might be hosting an exomoon. WASP 69b is also one of the exoplanet systems that will be observed as part of JWST cycle-1 GTO. This makes the system an excellent target to observe and follow up. We observed the system for 32 hrs at 150 MHz and 218 MHz using the upgraded Giant Metrewave Radio Telescope (uGMRT). Though we do not detect radio emission from the systems,  we place strong  $3\sigma$  upper limits of 3.3 mJy at 150 MHz and 0.9 mJy at 218 MHz. We then use these upper limits to estimate the maximum mass loss from the exomoon candidate. 
\end{abstract}


\begin{keywords}
radio continuum: planetary systems --- planets and satellites: magnetic fields --- planets and satellites: aurorae --- exoplanets
\end{keywords}

%%%%%%%%%%%%%%%%%%%%%%%%%%%%%%%%%%%%%%%%%%%%%%%%%%

%%%%%%%%%%%%%%%%% BODY OF PAPER %%%%%%%%%%%%%%%%%%
\section{Introduction} \label{Intro}
%%%%%%%%%%%%%%%%%%%%%%%%%%%%%%%%%%%%%%%%%%%%%%%%

The discovery of an exomoon is the next natural step in the galactic hierarchy of celestial objects. Signatures of alkali metals such as Na and K have been reported in the high-resolution spectra of about 20 transiting giant exoplanets \citep[e.g.,][]{Charbonneau, Wyttenbach}. By analogy with the Na escape signature from the Jupiter-Io system, \cite{Oza19} proposed that ionizing alkali clouds fueled by evaporative mass loss from exomoons orbiting these exoplanets can explain the observed alkaline exospheres of transiting exoplanet systems  \citep{Gebek20, Wyttenbach}.

The finding by \cite{Cassidy09} that exomoons around close-in gas giant exoplanets have stable orbits over astronomical timescales led to the demonstration that the stellar tide will melt the interiors of exomoons and evaporate their surfaces. In our solar system, due to tidal heating, Io exhibits evaporation in the form of extreme mass loss $\sim$ 1000 kg/s. While Io’s eccentricity and tidal heating are due to Europa and Ganymede locking it into a Laplace resonance (4:2:1) around Jupiter’s gravitational well \citep{Peale}, exomoons of close-in exoplanets have a large periodic eccentricity due to stellar forcing \citep{Cassidy09}. The impact of the stellar tide on the tidal heating rate ($\dot{E}_s$) is a strong inverse function of the planet's orbital period ($\tau_p$): $\dot{E}_s \propto \tau_p^{-5}$  \citep{Cassidy09}. The evaporation limit for an Io-mass exomoon orbiting a gas giant was found \citep{Oza19} to be within a critical orbital period
of $\tau_c$ = 1 day (for hydrodynamic mass loss, \citealt{2013MNRAS.433.2294P}) and $\tau_c$ = 2.6 days (for tidally driven mass loss, \citealt{2021arXiv210500917C}; but see \citealt{2021PASP..133i4401D}) consistent with the dearth of evaporated metals within the radius
resulting from this critical period. 

Transiting exoplanets are ideal candidates for exo-Ios because they allow for transit follow-ups with  JWST. However, an independent detection of these exomoons is necessary before considerable telescope time is devoted to investigating them. {One method of detecting exomoons is by observing the radio emissions due to (exo)planet-(exo)moon interaction. }

\begin{figure*}
\centering
\includegraphics[width=1\linewidth]{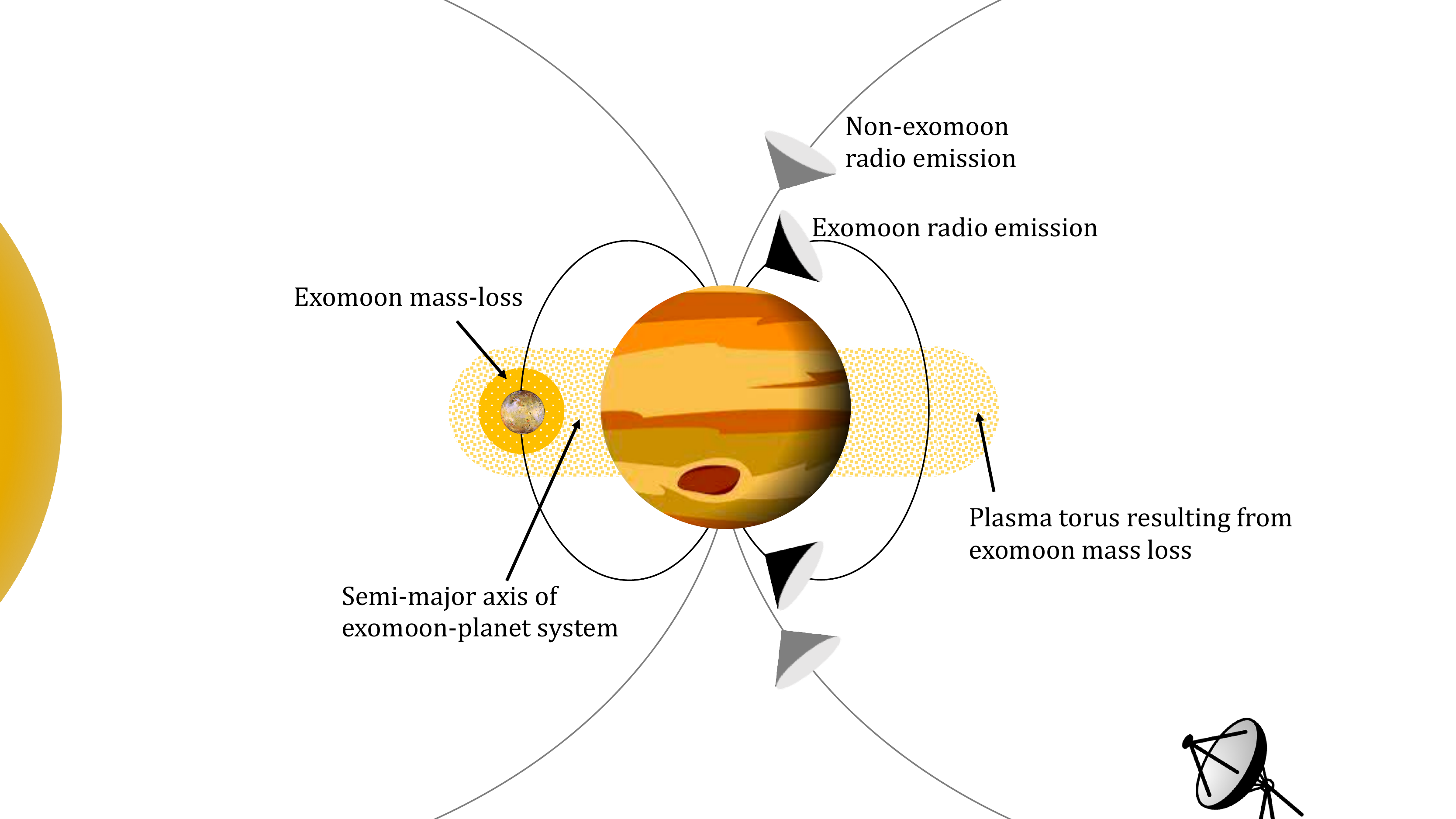}
\caption{A schematic representation of the ECMI emission process between an exoplanet, exomoon, and exoplasma-torus. The satellite semi-major axis $a_s$ is defined by dynamics and stability criteria \citep{Cassidy09}, the mass loss by stellar tides and irradiation \citep{Oza19}, and the scale height defines the volume of the plasma torus responsible for the beamed emission $S_{\nu_c}$. The cyclotron frequency $\nu_c$  = 2.8 $B_p$ determines the selected observational frequency and can result in non-exomoon emission (gray cone) and radio-loud exomoon-exoplanet emission (black cone) based on the field strength at the satellite orbit $B_s \sim$ $B_p$ ($R_p$/$a_s)^3$. }
\label{Fig1a}
\end{figure*}

{ The Io-controlled decametric (Io-DAM) emission  \citep{1964Natur.203.1008B} is one such example of planet-moon interaction that leads to a detectable signal. These emissions are generated by the interaction between Io and Jupiter's magnetic field. Io is a highly volcanic moon, and the volcanoes on its surface emit a large amount of ionized gas. Due to ongoing volcanism, the moon possesses an atmosphere of highly ionized SO$_2$ \citep{2007iag..book..231L}, which produces an ionosphere around the moon. This ionized gas is then captured by Jupiter's strong magnetic field, forming a plasma torus around the planet (see Figure \ref{Fig1a}). As Io orbits within this plasma torus, it generates a current that flows between the moon and the planet \citep{G69, Gris07}, giving rise to a unipolar inductor. The interaction between Io and the plasma torus also generates magnetic field oscillations known as Alfvén waves \citep{1987Sci...238..170B}, which lead to the production of electric fields parallel to the Jovian magnetic field line \citep{Neubauer80, C97, Saur04}. The electrons then accelerate along magnetic field lines, whose gyration produces radio emission via an electron cyclotron maser instability (ECMI) process \citep[e.g.][]{Wu79, Treumann06}. If a similar mechanism also operates in exomoon-exoplanet systems, then their emission might also be detectable, making these exomoons "Radio-loud", similar to Io. }

\begin{figure}
\centering
\includegraphics[width=1\linewidth]{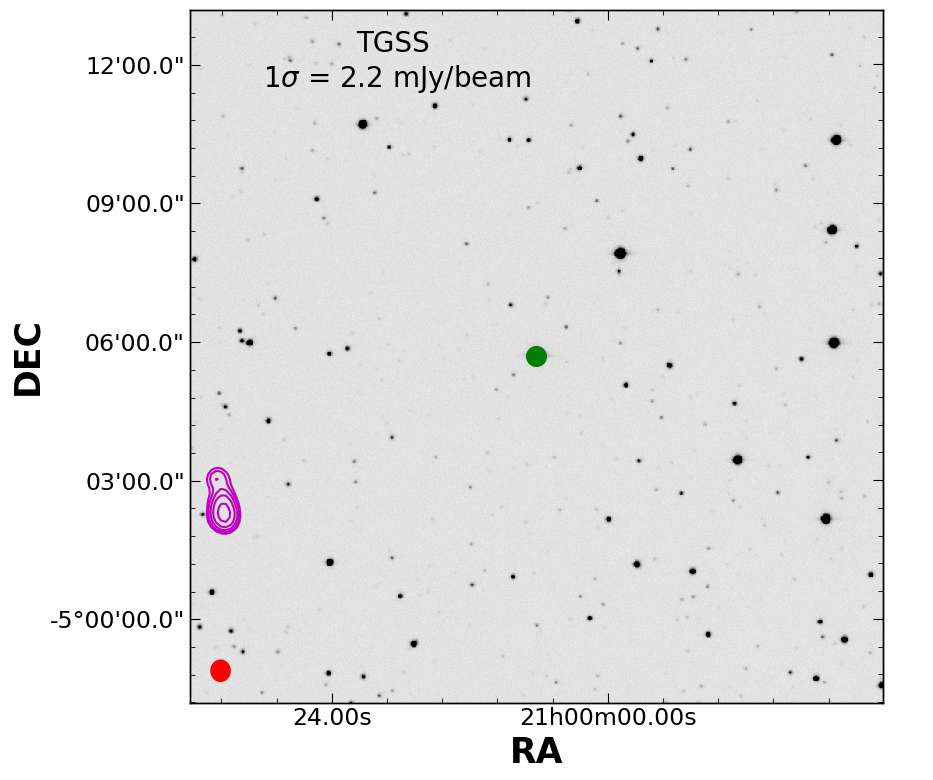}
\caption{The TGSS 150 MHz GMRT image (magenta contours) of the WASP-69 field at 150~MHz overlaid on the ztf g band image. The green circle marks the position of the WASP-69. The contours plotted are  5, 7, 10, 15, and 25~$\times\;\sigma$. The beam is shown as a red ellipse at the bottom left corner. }
\label{fig1b}
\end{figure}

\begin{figure*}
\centering
\includegraphics[width=0.5\linewidth]{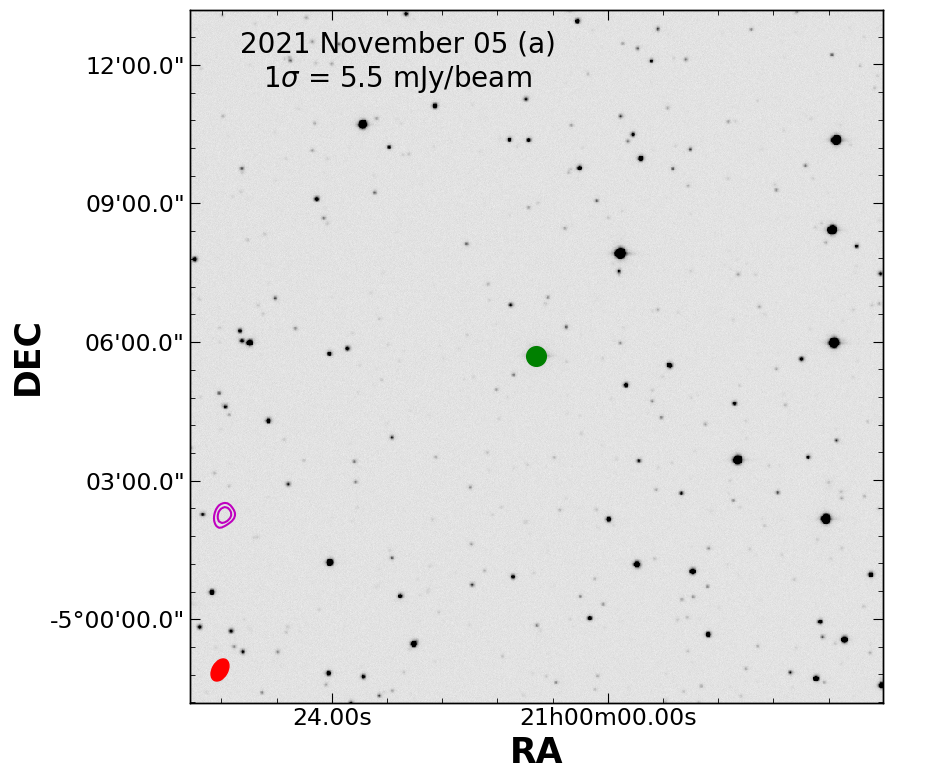}\includegraphics[width=0.5\linewidth]{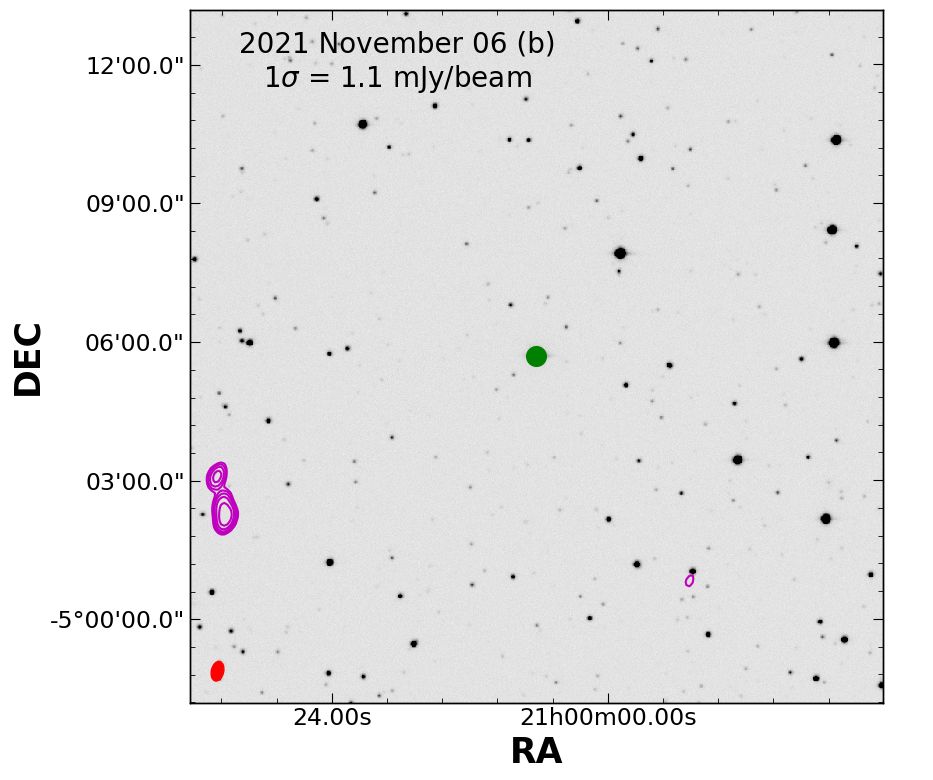}

\includegraphics[width=0.5\linewidth]{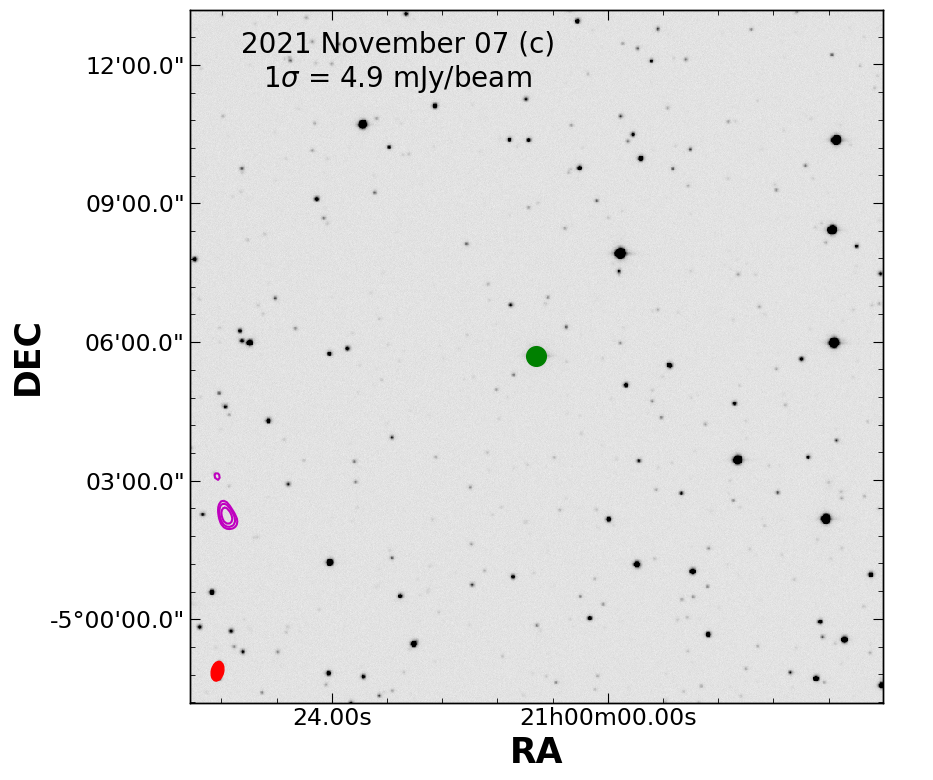}\includegraphics[width=0.5\linewidth]{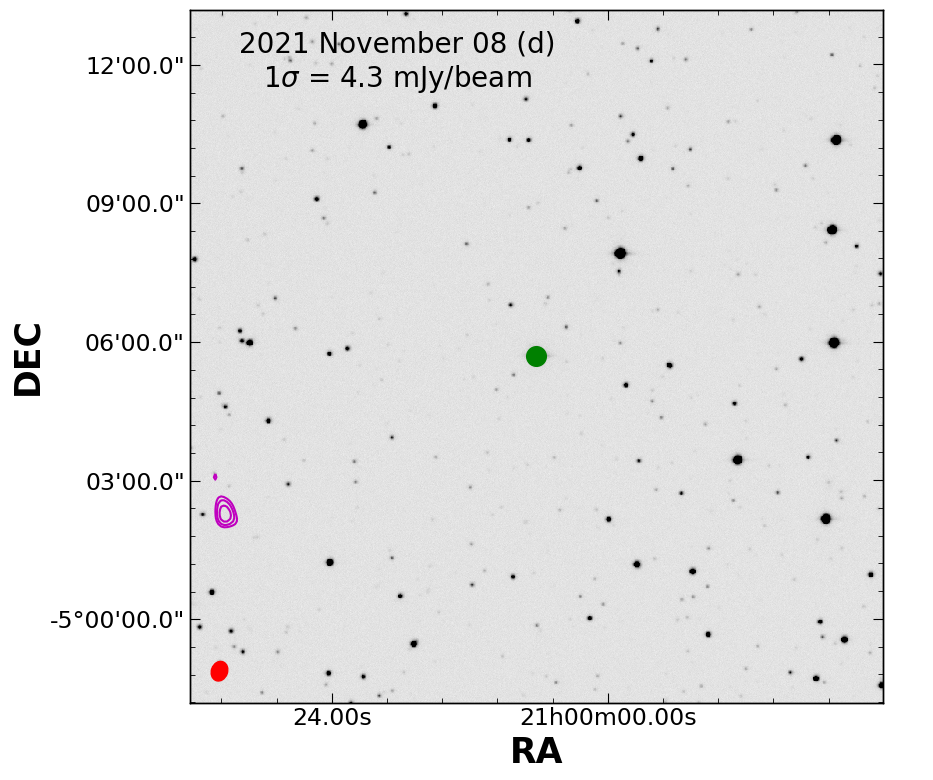}

\includegraphics[width=0.5\linewidth]{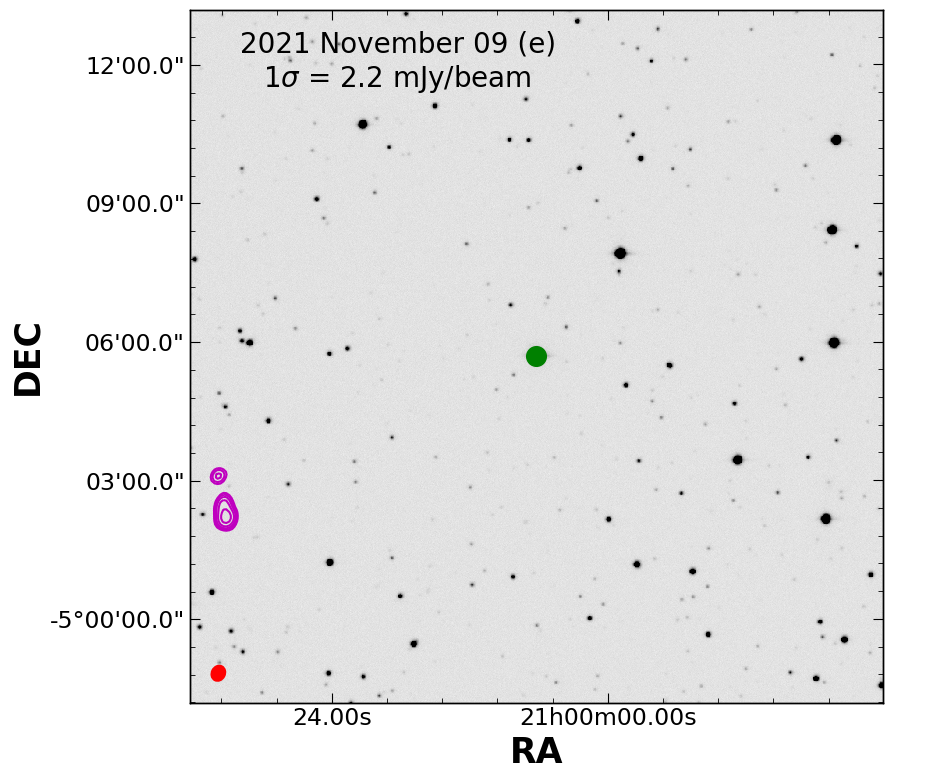}
\caption{The uGMRT image (magenta contours) of the WASP-69 field at 150~MHz  for each individual observation night overlaid on the ztf g band image. The green circle marks the position of the WASP-69. The contours plotted are  5, 7, 10, 15, and 25~$\times\;\sigma$. The beam is shown as a red ellipse at the bottom left corner. }
\label{fig1}
\end{figure*}

Several attempts have been made to detect radio emissions from exoplanets, but no successful detection has yet been reported \citep[see][]{Griessmeier17b,lazio17, Narang20, Narang21b, Turner20}. However, radio emission due to star-planet interaction from the star has been reported in a few cases \citep[e.g.,][]{Vedantham20, CC}. Recently \cite{2022arXiv221013298N} carried out the first dedicated search to detect radio emissions from a sample of exoplanets that showed possible signatures of an exomoon in their transmission spectra. Though no detections were made in that survey, it opened up the possibility of using radio observations to search for exomoons.  
Discovery of exomoons via radio emission hinges on several factors, including observing the emission at the correct frequency (which in turn depends on the magnetic field of the exoplanets, which are largely unknown), the emission being beamed towards us during our observation run, the evaporation from the exomoon being powerful enough to lead to a detectable signature. All these make the detection of radio emissions from planet-moon interactions a challenging but plausible experiment.

In this work, we present our observation of the hot-Saturn WASP-69b using the upgraded Giant Metrewave Radio Telescope (uGMRT) to study the planet-moon interaction and search for a volcanic exomoon. The star WASP-69 is a K5 star at a distance of  50 pc \citep{2021AJ....161..147B}. The WASP-69 system is host to WASP-69b, a hot-Saturn with a mass of 0.26 $\pm$ 0.017 $M_J$ (radius of 1.057 $\pm$ 0.047 $R_J$) and an orbital period of 3.86 days \citep[0.045 AU,][]{2014MNRAS.445.1114A}. A strong signature of Na was reported by  \cite{2017A&A...608A.135C}.  The atmosphere of WASP-69b has been studied widely with ground-based high-resolution spectroscopy and HST \citep{2018Sci...362.1388N, ES}. \cite{Oza19} attributed the Na detection in the atmosphere of WASP-69b to the presence of an exo-Io. 

Furthermore, the system WASP-69b will be observed as a cycle-1 JWST GTO target with both NIRCAM (proposal ID 1185 \citealt{2017jwst.prop.1185G}) and MIRI (proposal ID 1177 \citealt{2017jwst.prop.1177G}).  These data sets may possess infrared signatures of hot spots of a tidally-heated exomoon \citep{2013ApJ...769...98P}, which may be evident in NIRCAM and  MIRI observations. This makes WASP-69b not just an excellent candidate for uGMRT observations but also for follow-up ground and space-based observations. {WASP-69b was previously observed at 150 MHz with GMRT as part of the TIFR GMRT Sky Survey \citep[TGSS,][]{2017A&A...598A..78I}. The TGSS observations reached an rms of 2.2 mJy (see Figure \ref{fig1b}). However, no emission was detected from the source.}

In Section 2, we describe the details of the observations and the data reduction process. Next, we present our findings and discuss them in Section 3, followed by a summary in Section 4.

\begin{figure*}
\centering
\includegraphics[width=0.5\linewidth]{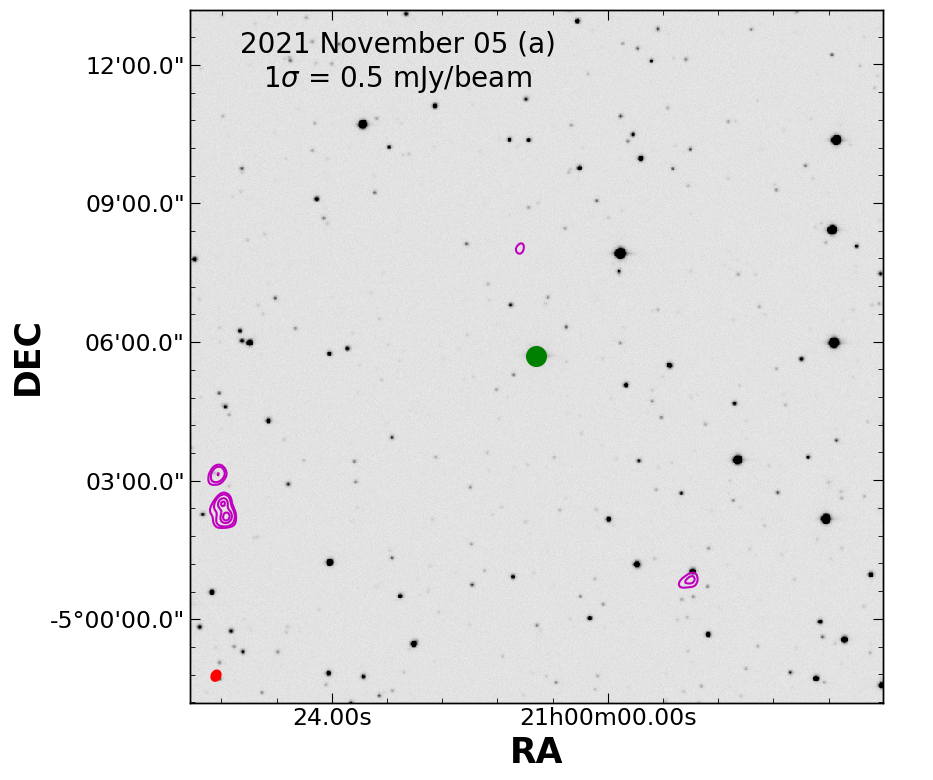}\includegraphics[width=0.5\linewidth]{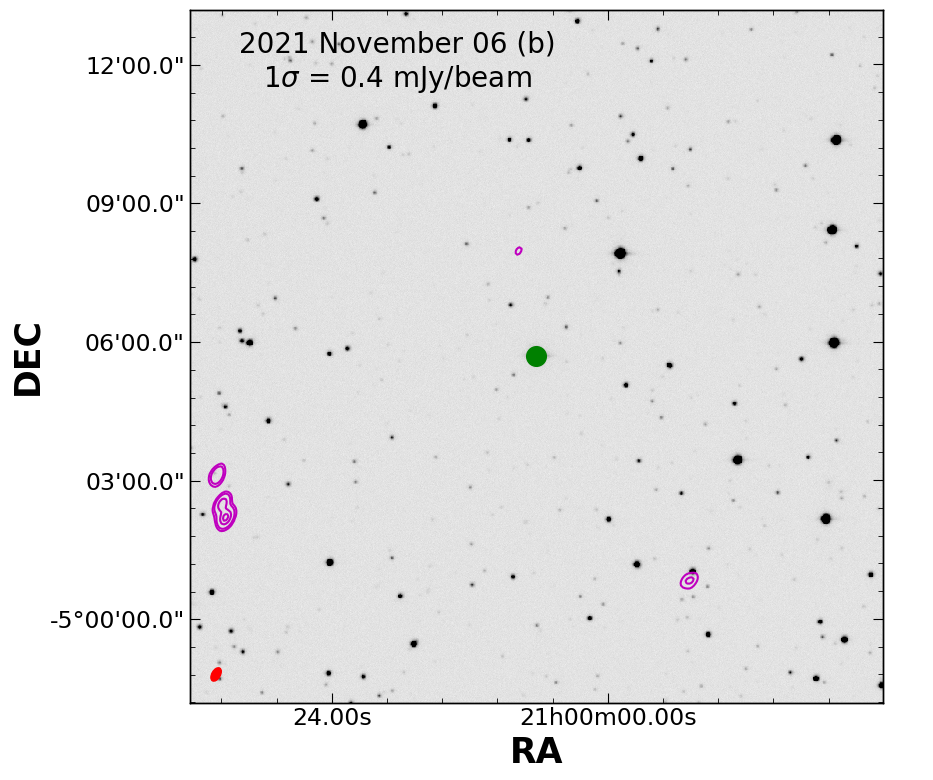}
\includegraphics[width=0.5\linewidth]
{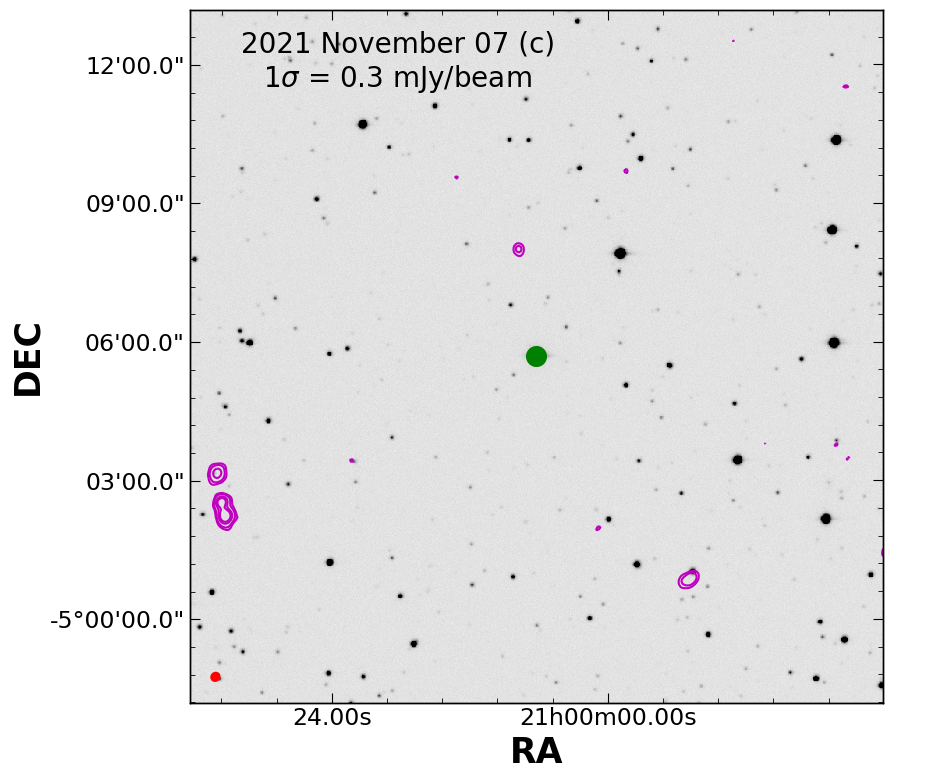}\includegraphics[width=0.5\linewidth]{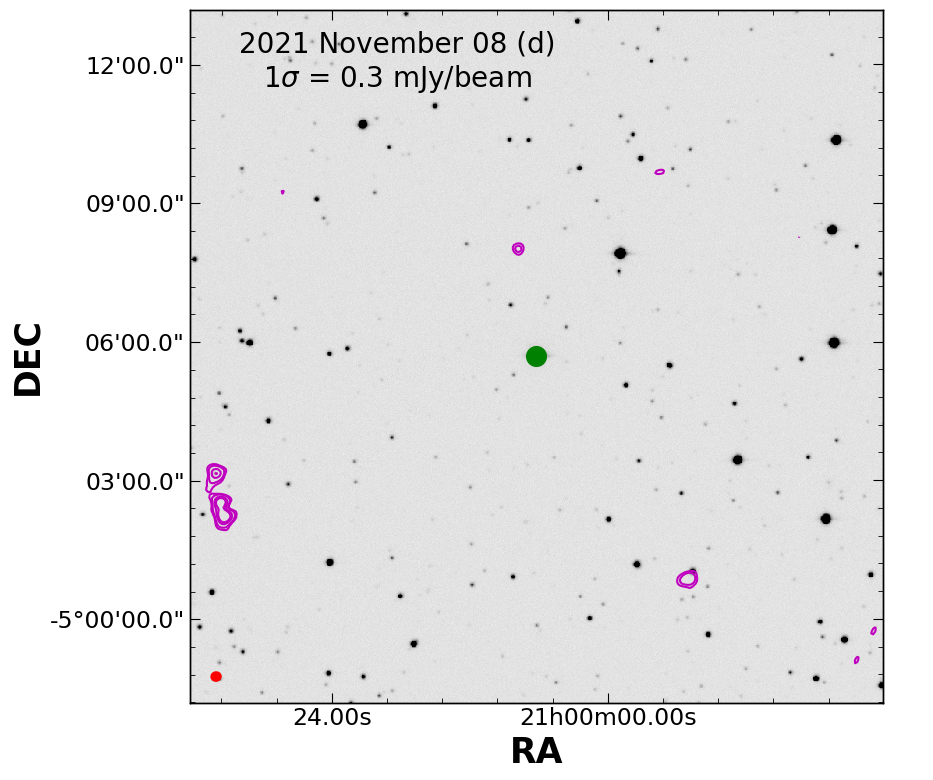}
\includegraphics[width=0.5\linewidth]{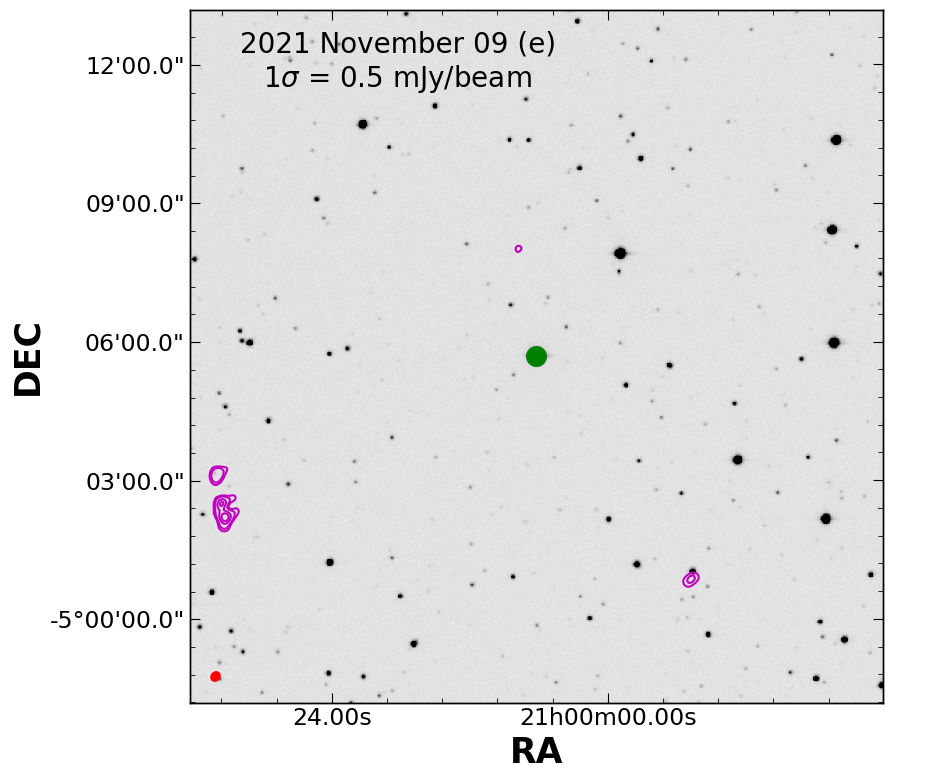}
\caption{The uGMRT image (magenta contours) of the WASP-69 field at 218~MHz  for each individual observation night overlaid on the ztf g band image. The green circle marks the position of the WASP-69. The contours plotted are  5,10, 30, and 50~$\times\;\sigma$. The beam is shown as a red ellipse at the bottom left corner. }
\label{fig2}
\end{figure*}

\section{Observation and data reduction}\label{observation}

The WASP-69 system was observed for 32 hrs with uGMRT (proposal ID 41\_068, PI Kaustubh Hakim). We observed the system in the band 2 (120-250 MHz) of uGMRT. The system was observed for five consecutive days. At each pointing, the system was observed for 6-7 hrs. The complete log of the observations is listed in Table \ref{Table1}. For all five pointings, the observation setup was the same. The flux calibrator 3C286 was observed at the beginning of the observation, while the flux calibrator 3C48  was observed at the end of the observation run. We observed the phase calibrator  2047-026 in a loop with the target WASP-69 with 27 mins of WASP-69 and 6 mins on the phase calibrator  2047-026.

\begin{table}
\centering
\begin{tabular}{ccccc}
\hline 
Date of  & Start time & Duration & rms  & rms  \\
observation & -- &--  &150 MHz & 218 MHz \\
-- & (UTC) & (hr) & (mJy/b) & (mJy/b)  \\ 
\hline
2021 Nov  05$^{th}$ & 0830 & 7 & 5.5 & 0.5 \\
2021 Nov  06$^{th}$ & 1030 & 7 & 1.1 & 0.4 \\
2021 Nov  07$^{th}$ & 0830 & 6 & 4.9 & 0.3 \\
2021 Nov  08$^{th}$ & 0830 & 6 & 4.3 & 0.3 \\
2021 Nov  09$^{th}$ & 0830 & 6 & 2.2 & 0.5  \\ 
\hline 
\end{tabular}%

\caption{Summary of observation and the rms sensitivity reached during our observation run}
\label{Table1}
\end{table}

To reduce the band 2 (120-250 MHz) uGMRT data, we used Source Peeling and Atmospheric Modeling (SPAM) pipeline \citep{Intema09b, Intema14b, Intema14}.  {SPAM is a python-based extension to Astronomical Image Processing System (AIPS) \cite{2003ASSL..285..109G}. SPAM was developed to reduce low-frequency radio interferometric observations using telescopes such as GMRT. SPAM has inbuilt routines for flagging RFI and bad data. SPAM also includes direction-dependent ionospheric calibration and image-plane ripple suppression, which can further improve the image quality. However, the SPAM pipeline does not support the processing of large fractional bandwidths ($\delta f/f>~0.2$). Thus natively, the SPAM is not capable of reducing the wideband data from uGMRT. A workaround for this is to split the bandwidth into smaller chunks (subbands) that can be processed independently. The calibrated output visibilities can then be jointly imaged to produce the final image. }

{The band-2 (120-250 MHz) of uGMRT has a break in the middle (165-185 MHz) and can be divided into two frequency ranges. Thus we decided to split band-2 into two different subbands with a bandwidth of about $\sim$ 30 MHz around regions of relatively low radio frequency interference.  We selected channel numbers from 600-1100 (500 channels) corresponding to a band center of $\sim$ 218 MHz and channel numbers from 1350-1750 (400 channels) corresponding to a band center of $\sim$ 150 MHz. These channels were relatively free of RFI. Therefore, we processed the two sub-bands independently and produced the final images. The rms noise of these two sub-bands is very different, so we decided not to combine the two images to produce a wideband image.}

\section{Results}\label{results}

\subsection{Observed upper limits on the radio flux density}

The WASP 69 system was observed with uGMRT for five days in band 2  (120-250 MHz), totaling 32 hrs.  In Figure \ref{fig1}, we show the band-2 150 MHz images of the WASP-69 field, while in Figure \ref{fig2}, we show the band-2 218 MHz images of the WASP-69 field. In Table \ref{Table1}, we have listed the rms reached during these observations.  The rms value for the WASP-69 field ranges from 1.1--5.5 mJy/beam at 150 MHz and between 0.3--0.5 mJy/beam at 218 MHz. Using the lowest value of rms and assuming 3$\times  rms$ as an upper limit to the radio flux density $S_{\nu}$, we get $S_{\nu} = 3.3$~mJy at 150 MHz and $S_{\nu} = 0.9$~mJy at 218 MHz. Our observations at 218 MHz are some of the deepest observations that have been carried out at these frequencies \citep[e.g.,][]{Etangs09, Etangs11, Narang20,gorman18,2022MNRAS.515.2015N}.

\subsection{Maximum radio power}

The maximum radio power $P_{\nu}$ emitted by the exoplanet-exomoon system can be calculated from the observed upper limits on the radio flux density \citep[e.g.,][]{Lazio04, Gris07}: $P_{\nu} = S_{\nu} \Delta \nu \, \Omega d^2$, where $\Delta \nu $ is the bandwidth of emission such that  $\Delta \nu  = \nu_c/2$, $\Omega=0.16$ \citep[average value of Io-DAM][]{Zarka04} is the angle of the emission cone, and $d$ is the distance of the exoplanet from Earth. We find the maximum radio power that could be emitted from the WASP-69 system is 9 $\times 10^{14}$ W (at 150 MHz) and 4 $\times 10^{14}$ W (at 218 MHz). Compared to the maximum radio power emitted from the Io Flux Tube (IFT) of $\sim$ 10$^{8}$--10$^{10}$ W \citep{2001AdSpR..27.1915B}, our upper limits are 10$^{4.9}$--10$^{6.9}$ times higher at 150~MHz and 10$^{4.6}$--10$^{6.6}$ times higher at 218~MHz. 
%The maximum exo-Io to Io power ratio is still larger than the tidal enhancement factor $\eta_{\mathcal{T}} \sim$ 10$^{5.3}$, predicted for an Io-sized satellite tidally-heated to 1296K (Dobos \& Turner 2013; \citep{Oza19}. For certain exoplanets, magnetic star-planet interactions were predicted to yield a net power of 10$^{19}$ Watts \citep{Saur13}. Here, regardless of the physical mechanism, our observations rule out star-planet interactions of this magnitude. 

\subsection{Constraining the mass loss rate from the exomoon}

The maximum emitted radio power from an exoplanet-exomoon interaction depends on the plasma mass density and the magnetic field \citep{Neubauer80}. At low plasma densities, the radio power $P_{-}$ scales linearly with the magnetic field at the satellite location, $B_s$, and as the square root of plasma mass density $\rho_s$. At high plasma densities, the radio power $P_{+}$ scales with the square of the magnetic field and is independent of plasma mass density \citep{Noyola14},

\begin{align}
 \label{eq2}
 P_{-} & \propto B_s \sqrt{\rho_s}, \\
 P_{+} & \propto B_s^2 .
 \label{eq3}
\end{align}

Although the determination of the magnitude of plasma density is beyond the scope of this paper, these two limits can be used to make qualitative arguments on the mass loss rate from the exomoon $\dot{M}$, which scales linearly with plasma density: $\dot{M} \propto \rho_s$. Moreover, for a given planetary magnetic field $B_0$, the magnetic field at the satellite location drops as $B_s \propto B_0 / a_s^3$ (see Sect.~\ref{sec:disMagField} for discussion on magnetic field strength). Therefore, at the lower limit of plasma density, the mass loss rate from the exomoon is proportional to the sixth power of the exomoon semi-major axis,

\begin{align}
 \label{eq4}
 \dot{M} \propto (\frac{P_{-}}{B_s})^2 =  (\frac{P_{-}}{B_0})^2 \, \, a_s^6.
\end{align}

\noindent
For stable exomoons, the value of $a_s$ is between the Roche limit and half the hill radius \citep{Cassidy09}. For WASP 69 b, this range for a satellite having a composition similar to Io is between 1.16--2.1 $R_J$ \citep{Oza19}. This is much closer than the orbital distance of 5.9 $R_J$  of Io around Jupiter. Taking  $a_s = 1.63 R_J$ (average of the Roche limit and half the Hill radius), the minimum radio power ratio of WASP 69 b to IFT of 10$^{4.9}$ at 150~MHz, and the magnetic field ratio of $\sim13$ ($B_{0, W69b} = 54$~G at 150~MHz, which is fixed by the search frequency, $\nu = 2.8 B_0$, cf. $B_{0, Jup} = 4.17$~G)), we find that the mass loss rate from a hidden exo-Io is roughly 17000 times higher than Io. However, to better constrain the mass loss from the system, further observations of the companion alkali doublet K predicted to be roughly $\sim$10 $\times$ less abundant than Na \cite{Gebek20}, are required in ground-based high-resolution spectroscopy. With NIRCAM/MIRI (proposal ID 1185 and 1177) and JWST, volcanically-vented molecules such as SO$_2$ CO$_2$, CO are expected to be apparent due to tidally-heated volcanism as seen at Io from both ground-based \cite[CRIRES/VLT][]{2015Icar..253...99L} and space-based spectrographs \citep[JIRAM/JUNO,][]{2020JGRE..12506508M}.

\section{Discussion}

There could be several reasons why no radio emission was detected from these systems. In the following subsection, we discuss some of them.

\subsection{Cyclotron Frequency and Gas Giant Magnetic Field Strength} \label{sec:disMagField}

The cyclotron frequency $\nu_c$ for emission is given as $\nu_c = 2.8 B_0$ where $B_0$ is in Gauss and $\nu_c$ in MHz. The choice of observing the system in band-2 (120-250 MHz) of uGMRT was based on hot Saturns having magnetic fields of $\sim$ 40-100 G \citep{Yadav17}. If the magnetic field is not within this range, we would not detect it. Furthermore, the probable detection of radio emission from the hot Jupiter $\tau$ Bo\"otis b by  \cite{Turner20} between 15-30 MHz has challenged the notion of hot-Jupiters and hot-Saturns having a strong magnetic field. \cite{Turner20} estimate the magnetic field of  $\tau$ Bo\"otis b to be in the range of 5–11 G. If WASP-69b also possesses such a small magnetic field, then no emission would be detectable in the band-2 (120-250 MHz) of uGMRT. To test this hypothesis, we further computed the magnetic field of WASP-69b.  We followed the formalism from \cite{Yadav17} to estimate the magnetic fields. We used the evolution models from \cite{Thorngren2018} to derive the heat flux from the interiors of the planets \citep[also see][]{Christensen2009}. The magnetic field on the dynamo surface is given as \citep[from][]{Reiners2010}

\begin{equation}
    B_{rms}^{dyn} \, \mathrm{[G]} = 4.8\times10^3 (M_P L_P^2)^{1/6}  R_P^{-7/6} .
\end{equation}

\noindent
where $M_P$, $L_P$, and  $R_P$ are the mass, luminosity, and radius of the planet (all normalized to solar values). Assuming  scaling law for the dynamo radius from \cite{Yadav17}, the dipole magnetic field strength at the pole is thus

\begin{equation}
    B_{dipole}^{polar} = \frac{B_{rms}^{dyn}}{\sqrt{2}} \left(\frac{R_{dyn}}{R_P} \right)^3
\end{equation}

\noindent
{where $R_{dyn}$ is the dynamo radius.} By plugging in the values for the WASP 69 system, we estimate the $B_{dipole}^{polar} = 15$ G. This gives $\nu_c = 42$ MHz. This is much lower than the frequency at which we observed the system.  

\subsection{Time variable emission}
The decameter emission from Jupiter due to the interaction between Jupiter and Io or Jupiter and the solar wind is modulated with a period of a few milliseconds to months \citep{Lecacheux04, Marques17,1996GeoRL..23..125Z,2014A&A...568A..53R}. The emission from exo-moons can also be highly time variable and modulated with the moon's phase around the planet. The emission from these exomoons can be highly beamed \citep[e.g.,][]{1998JGR...10326649Q, Zarka04,2022JGRA..12730160L} and emitted in a narrow cone. In such a case, the emission will only be observable during certain phases of the moon around the planet and the planet around its host star. During our observation run, we cover about 35 $\%$ (32 hrs / 92.6 hrs) of the orbital phase of the planet. However, if the emission cone was not pointed towards Earth, we would miss it.

\subsection{Exomoon Flux Density}
As stated in \cite{2022arXiv221013298N}, radio emission due to exomoon-exoplanet interaction can be inherently weak. If the mass loss rate from the exomoon is lower, then we will not be able to detect any emission. Furthermore, WASP 69 is located at 50 pc, and with our current sensitivity of telescopes, we will not be able to detect any signal if the strength of the radio emission from the system is at the same level as Io Flux Tube. The next generation of telescopes with high sensitivity are perhaps required for the detection of radio emission arising from exoplanet-exomoon interaction.

\section{Summary}
This work presents the first radio observations of the exoplanetary system WASP-69 using uGMRT.  The WASP-69 system is an exo-Io candidate system based on the presence of strong alkaline metal lines in the transmission spectra of the planet \citep{Oza19}. The WASP-69 system was observed in  band-2 (120-250 MHz) of uGMRT. For this analysis, we divided band-2 of uGMRT into two sub-bands at 150 MHz and 218 MHz. We observed the WASP-69 field for 32 hrs covering about 20\% of the orbital phase of the planet over five pointings.  At 150 MHz, we achieved a 3 $\sigma$ upper limit of 3.3 mJy, while an upper limit of 0.9 mJy was obtained at 218 MHz. However, no radio emission was detected from the system.  This implies that the exomoon may not be radio-loud at the synchrotron frequencies searched for due to either a lower mass loss rate from the exomoon or a lower magnetic field strength of the parent planet. Moreover, the emission could be highly time variable and beamed. Therefore,  deeper and more frequent observations at lower radio frequencies of the systems than what is currently possible are thus required to detect exomoons.

{The upcoming generation of radio telescopes, including the next-generation VLA (ngVLA) \citep{2019BAAS...51g..81M} and Square Kilometre Array (SKA)  \citep{2009IEEEP..97.1482D}, will surpass the current telescopes in sensitivity. This increased sensitivity creates the possibility of detecting faint signals originating from the interaction between exoplanets and exomoons even at a much lower frequency. The magnetic field strength for WASP 69b is estimated to be 15 G, which is similar to the estimates derived for other exoplanets \citep[e.g.,][]{Yadav17,2022arXiv221013298N}. The emission generated from the exoplanet-exomoon interaction from exoplanets with such low magnetic field strengths will fall below 100 MHz in frequency. Consequently, the SKA appears to be a well-suited instrument for detecting these elusive exomoons.}

\section{Acknowledgment}

This work is based on observations made with the Giant Metrewave Radio Telescope, which is operated by the NCRA TIFR and is located at Khodad, Maharashtra, India.  KH is supported by the FED-tWIN research program STELLA funded by the Belgian Science Policy Office (BELSPO).

\section{Data availability}

The data presented in this article are available on the GMRT archive at https://naps.ncra.tifr.res.in/goa/, and can be accessed with proposal id $ 41\_068$.

\bibliographystyle{mnras}

\bibliography{Exo}
\bsp	% typesetting comment
\label{lastpage}
\end{document}